# AI-Native Integrated Sensing and Communications for Self-Organizing Wireless Networks: Architectures, Learning Paradigms, and System-Level Design

S. Zhang, M. Feizarefi, A. F. Mirzaei

Department of Electrical Engineering,
New Jersey Institute of Technology, Newark, NJ 07102, USA.
Corresponding author: S. Zhang, email: {saiminzhang@njit.edu}

**ABSTRACTS:**
Integrated Sensing and Communications (ISAC) is emerging as a foundational paradigm for next-generation wireless networks, enabling communication infrastructures to simultaneously support data transmission and environment sensing. By tightly coupling radio sensing with communication functions, ISAC unlocks new capabilities for situational awareness, localization, tracking, and network adaptation. At the same time, the increasing scale, heterogeneity, and dynamics of future wireless systems demand self-organizing network intelligence capable of autonomously managing resources, topology, and services. Artificial intelligence (AI), particularly learning-driven and data-centric methods, has become a key enabler for realizing this vision. This survey provides a comprehensive and system-level review of AI-native ISAC-enabled self-organizing wireless networks. We develop a unified taxonomy that spans: (i) ISAC signal models and sensing modalities, (ii) network state abstraction and perception from sensing-aware radio data, (iii) learning-driven self-organization mechanisms for resource allocation, topology control, and mobility management, and (iv) cross-layer architectures integrating sensing, communication, and network intelligence. We further examine emerging learning paradigms, including deep reinforcement learning, graph-based learning, multi-agent coordination, and federated intelligence that enable autonomous adaptation under uncertainty, mobility, and partial observability. Practical considerations such as sensing-communication trade-offs, scalability, latency, reliability, and security are discussed alongside representative evaluation methodologies and performance metrics. Finally, we identify key open challenges and future research directions toward deployable, trustworthy, and scalable AI-native ISAC systems for 6G and beyond.

## 1. Introduction

Future 6G wireless networks are envisioned to be AI-native and multifunctional, not only connecting devices but also perceiving and reacting to their environment in real time [1]. An emerging hallmark of 6G is Integrated Sensing and Communications (ISAC), which embeds radar-like sensing capabilities into the communication infrastructure [2]. In parallel, network management is shifting toward self-organizing principles, where AI-driven control loops enable autonomous topology adaptation, resource optimization, and policy self-tuning with minimal human intervention [3], [4]. The fusion of ISAC with AI-driven self-organization promises a new class of perceptive intelligent networks that sense their surroundings and learn to optimize operations accordingly.

Integrating sensing into the wireless network addresses both technical and economic drivers. On one hand, joint sensing-communication functionality can dramatically improve spectral and energy efficiency by using a single waveform and hardware platform for dual purposes [5]. This reduces cost and overhead compared to separate radar and communication systems [6]. Indeed, the International Telecommunication Union (ITU) has identified ISAC as a key usage scenario for 6G due to its potential in real-time environment reconstruction, localization, and situational awareness [7]. On the other hand, the rich sensory data collected by ISAC networks (e.g. echo signals, channel state information) can serve as an enabler for machine learning, allowing the network to "see" and build environmental intelligence. For instance, future networks are expected to use sensed context (user location, objects, blockages) to proactively adjust beams, routes, and resource allocation decisions. This deeper integration blurs the boundary between communication and sensing: communications can assist sensing (e.g. multiple nodes cooperatively sensing a target), and sensing can assist communications (e.g. radar-derived location aiding beam alignment) [8], [9].

Meanwhile, AI-native control is becoming indispensable to manage the complexity of next-generation networks. 6G networks will be far more dense, heterogeneous, and dynamic, featuring millimeter-wave/Terahertz links, massive antennas, intelligent surfaces, aerial relays, and IoT nodes with sporadic mobility. Traditional static optimization and rule-based management cannot adapt quickly enough to rapid channel variations and network reconfigurations [10]. Thus, researchers and standardization bodies (e.g. 3GPP Release 18 and beyond) are incorporating machine learning (ML) and closed-loop automation into RAN architecture. Unlike legacy Self-Organizing Network (SON) approaches that relied on pre-programmed heuristics, 6G aims for networks that learn and self-evolve, performing self-optimization, self-healing, and self-configuration by continually analyzing data and improving decisions [11]. Early signs of this paradigm shift include 3GPP's studies on AI/ML frameworks for RAN (for lifecycle management, inference monitoring, and fallback mechanisms) and industry initiatives like the O-RAN Alliance's RIC (RAN Intelligent Controller) for hosting AI-driven xApps.

In summary, an AI-native ISAC network is a vision of a wireless system that simultaneously communicates and senses, and uses the sensed information to autonomously organize and optimize itself. This survey provides a comprehensive review of the state of the art in this interdisciplinary domain. We first introduce a taxonomy of ISAC systems and AI-based self-organization approaches (Section II). Then we delve into core technical areas: (1) ISAC techniques including signal models, sensing modalities, waveform design, and sensing-communication co-design (Section III); (2) AI-based self-organizing network control, covering learning paradigms for topology control, mobility management, resource allocation, etc. (Section IV); and (3) the intersection of ISAC and AI-native control, highlighting how sensing feedback loops can drive self-optimizing decisions (Section V). We also survey evaluation methodologies for these systems (Section VI), discuss open challenges (Section VII), such as performance trade-offs, data and privacy issues, reliability, and standardization, and conclude with future outlook (Section VIII). Our focus is on recent advances (2020–2024) in top-tier journals and conferences (e.g. IEEE TWC, JSAC, ComMag, Sensors), including emerging ideas like joint radar-communication waveforms, deep reinforcement learning (DRL) for dynamic network control, graph neural networks (GNNs) for networking problems, and federated/self-supervised learning for distributed intelligence. The goal is to distill key architectural insights, learning paradigms, and system-level design principles that will shape 6G self-organizing networks with integrated sensing and communications.

## 2. Taxonomy of ISAC and AI-Native Self-Organizing Networks

To organize this fast-evolving field, we present a taxonomy from two perspectives: ISAC system architectures and AI-enabled network control strategies. Figure 1 depicts one facet of ISAC architectures, classifying systems by radar deployment and target cooperation. In general, ISAC transceivers can operate in monostatic mode (transmitter and receiver co-located) or bistatic/multi-static mode (separate transmitting and receiving nodes), and targets may be device-based (cooperative devices transmitting signals) or device-free (passive targets reflecting signals) [12]. These distinctions yield four typical configurations: (a) monostatic ISAC with device-based targets, (b) monostatic with device-free targets, (c) bistatic with device-based targets, and (d) bistatic with device-free targets. Each has implications on interference, synchronization, and sensing performance, as discussed shortly. On the other hand, AI-driven network control can be categorized by learning paradigm (e.g. supervised vs. reinforcement learning, centralized vs. federated learning) and by the network function being automated (topology management, mobility control, resource allocation, etc.). We will classify major contributions along these lines. Table 1 summarizes key dimensions of the survey's scope, mapping core technology areas to representative techniques and references.

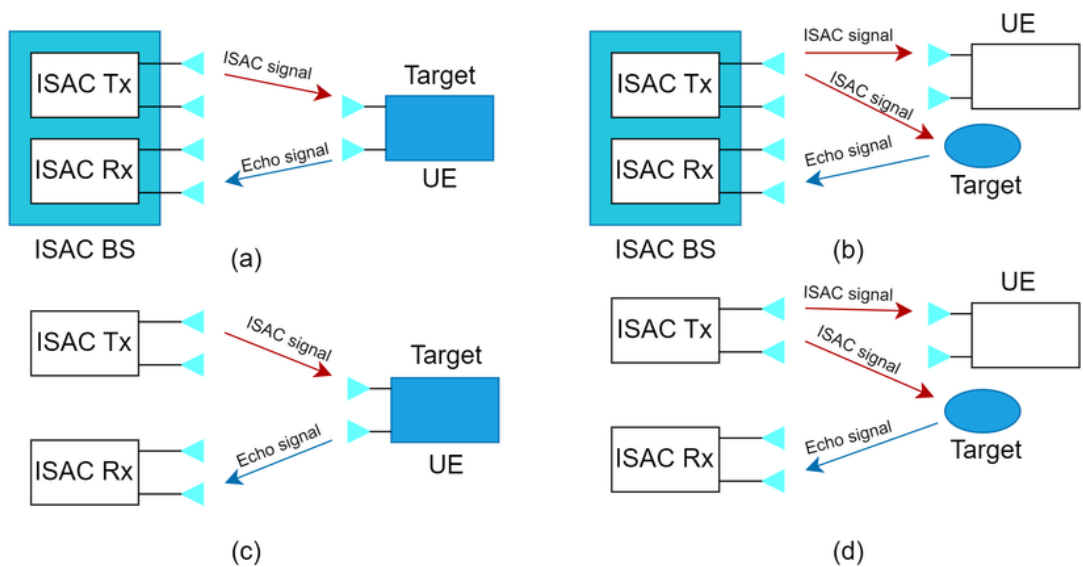

Figure 1: Taxonomy of ISAC system configurations (communication-sensing integration architectures). (a) Monostatic ISAC, device-based target: a single base station (BS) both communicates with and senses a cooperative user device (the user assists sensing by sending reference signals). (b) Monostatic ISAC, device-free target: a BS senses a passive object (no signal feedback) while simultaneously communicating with other users; self-interference from the BS's transmitter to its receiver is a challenge here [14], [15]. (c) Bistatic ISAC, device-based: separate TX and RX (e.g. two BSs or a BS and an access point) cooperate, where the target is a device that responds to probing, this eliminates self-interference at the receiver and can improve SNR [16]. (d) Bistatic ISAC, device-free: separate TX/RX sense a non-cooperative object; this requires coordination and backhaul between nodes but allows flexible geometry for sensing. These configurations illustrate how ISAC can be deployed in both cellular (monostatic) and distributed (multi-static) network settings.

ISAC Architectures: In any ISAC system, a fundamental design choice is how tightly the sensing and communication functions are integrated. In coexistence architectures, radar and communication share spectrum or hardware but are managed largely independently (with interference mitigation). In dual-functional or fully integrated architectures, a single waveform and signaling scheme performs both functions. Our taxonomy focuses on the latter. The monostatic ISAC case (Figure 1a, 1b) is analogous to a cellular base station using its downlink/uplink signals for radar sensing of the environment. It benefits from simpler hardware (one site) and easy synchronization between TX and RX [18], but suffers from self-interference if transmit and receive operations overlap. Monostatic ISAC is feasible in a time-division manner (transmit then receive echo) or with advanced self-interference cancellation to enable simultaneous transmit-receive. In contrast, bistatic/multi-static ISAC (Figure 1c, 1d) involves geographically separated transmitters and receivers[16]. This eliminates self-interference and allows more spatial diversity, for example, multiple BSs can jointly illuminate a target and share received echoes – improving detection coverage and robustness to blockages [19]. However, it requires tight time/frequency synchronization and real-time information exchange (e.g. sharing waveforms or sensing results) over backhaul links [20]. Multi-static setups (multiple TX/RX nodes) effectively turn the network into a distributed sensing array, reducing target occlusion and enabling cooperative sensing. A current trend in 6G is indeed toward networked sensing, where many infrastructure nodes (cell sites, vehicles, IoT sensors) collaborate to sense the environment, marking a shift from traditional monostatic radars to ubiquitous sensing architectures [21].

Another axis of the taxonomy is target involvement. In device-based sensing, the object to be sensed is an active device in the network (e.g. a user equipment (UE) that can transmit reference signals). This is akin to radio-based positioning or two-way ranging: the network and device cooperate to measure the round-trip time, angle, etc. Device-based ISAC can achieve high localization accuracy since the device can report observations and assist the process [13]. In device-free sensing, by contrast, the targets are ambient objects or persons not actively emitting known signals (e.g. detecting a pedestrian or drone via their radar cross-section). This is more challenging; the network must discern reflections from unknown objects amidst clutter. The advantage is providing sensing services (like intrusion detection, vehicular traffic monitoring, remote sensing) without requiring every object to carry a radio [22], [23]. Device-free ISAC relies on advanced signal processing and often multiple viewpoints (multi-static) to detect and localize targets. Many 6G use cases fall in this category: for example, a 6G BS network could detect and track non-cooperative drones for security [24] or monitor vehicles on a smart road for collision avoidance. We note that device-free sensing can also leverage existing users' signals as illuminators (known as passive sensing or passive radar, where the network listens for reflections of communication signals).

AI-Native Control Paradigms: In parallel with the physical architectures above, our taxonomy considers the AI techniques empowering network self-organization. Broadly, AI paradigms in this context include: (i) Supervised learning, used for tasks where labeled data is available to train models (e.g. using sensed channel state information to predict optimal configurations [25]). (ii) Unsupervised and self-supervised learning, applied when explicit labels are unavailable, for instance, clustering nodes or extracting features from high-dimensional sensing data for anomaly detection. These helps exploit raw sensory data and have been used in ISAC for target classification and recognition with limited labeled samples [26]. (iii) Reinforcement learning (RL), where an agent (e.g. a base station or distributed agent in each node) learns to take actions (adjust beams, reposition nodes, allocate resources) through trial-and-error to maximize long-term rewards [27], [28]. RL is natural for self-optimizing networks because it does not require a priori labels, the agent learns from the environment feedback. Variants like multi-agent RL (for cooperative decisions among multiple nodes) and deep RL (using deep neural networks as function approximators) have shown success in dynamic spectrum management, handover optimization, and UAV control in recent studies [28], [29]. (iv) Graph-based learning, which leverages graph neural networks or graph algorithms to handle the relational structure of network nodes [30]. GNNs can generalize resource allocation or routing decisions to varying network topologies, exploiting the underlying wireless network graph to improve scalability. (v) Federated and distributed learning, which train models across multiple nodes' data without centralizing it [31]. This is crucial when raw sensing data is privacy-sensitive or too voluminous to transmit; instead, local models are trained at edge devices (e.g. vehicles, BSs) and only model updates are aggregated [32]. Federated learning has been explored for distributed sensing (e.g. collaborative radar signal classification [33]) and for communication tasks like beam prediction across vehicles [34].

These AI paradigms intersect with various self-functions in the network. We identify key self-organizing functions: self-configuration & topology control (e.g. deciding the network graph, which links or nodes are active, how to form clusters or cell associations), self-optimization of resources (dynamic allocation of spectrum, power, time slots, antenna beams to optimize throughput or latency), self-healing (automatic detection and mitigation of faults or coverage holes, possibly using sensed data to reroute around failures), and policy adaptation (tuning protocol parameters or MAC/routing policies on the fly via learning). Each of these can be mapped to AI solutions: for instance, deep RL for dynamic spectrum access, GNN-based optimization for power control, or transfer learning to adapt scheduling policies to new environments. Table 1 outlines how different learning techniques map to these control tasks, with examples from recent literature.

Table 1: Key Domains and Techniques in AI-Native ISAC Networks

| Aspect | Example Techniques | Representative Recent Work (2020–24) |
|---|---|---|
| ISAC Waveform Design | Dual-functional radar-comm waveforms; OFDM and chirp fusion; adaptive waveform synthesis (learning-based) | Liu et al. (JSAC'22) [5]; Jiang et al. (TVT'25) – DL-based ISAC waveform optimization [35]. |
| Sensing-Comm Co-Design | Power/time allocation between sensing vs. data; interference mitigation techniques; Pareto-optimal trade-offs | Zhang et al. (TSP'21), rate vs. detection trade-off[36]; Kumari et al. (Sensors'20), waveform agility for comm-assisted radar. |
| Topology & Mobility Control | Multi-agent RL for dynamic cell planning; UAV placement optimization (learn to reposition aerial base stations); graph algorithms for relay selection | Wang & Farooq (MASS'23) – DRL for 3D UAV deployment[37][38]; Chen et al. (ICC Workshops'23) – GNN for relay/power optimization [30]. |
| Resource | Deep learning approximations of scheduling and power control; GNN- | Eisen et al. (TWC'20) – NN for power control (interference graph); Chen et al. |

| Aspect | Example Techniques | Representative Recent Work (2020–24) |
|---|---|---|
| Allocation | based power control; distributed Q-learning for channel access | (ICC'23) – FD-GNN for full-duplex power allocation [39]. |
| Beam Management | Sensing-aided beam selection (radar/LiDAR+camera to predict best beam); deep RL for beam sweeping policies; codebook design via learning | Charan et al. (JSAC'22) – vision-aided mmWave beam prediction; Nie et al. (ArXiv'23) – multimodal (RGB + radar) beam tracking[9]; (also see Section V). |
| Self-Healing & Fault Mgmt | Anomaly detection via network sensing (e.g. detect jamming or failures); federated learning for intrusion or fault pattern recognition | Liu et al. (Comm. Surveys'21) – SON with ML for fault management; Bhardwaj & Kim (ICUFN'23) – federated learning for beam failure detection [40]. |
| Policy/Protocol Adaptation | Meta-learning for quick reconfiguration; AI-tuned handover and routing parameters; cooperative multi-agent policies (e.g. for load balancing) | Foukas et al. (IEEE Network'2020) – AI for slicing adaptation; Lee et al. (Access'22) – deep Q-learning for handover optimization in dense networks. |
| Distributed Learning & Privacy | Federated learning for model training on distributed sensing data; incentive mechanisms for data sharing; privacy-preserving learning (differential privacy, etc.) | Saputra et al. (T-ITS'21) – FL for vehicular sensing; Das et al. (IJIN'25) – survey of AI in 6G including FL, privacy [32], [41]. |

Table 1 highlights that learning-based approaches now permeate all layers of network design, from physical-layer waveform adaptation to MAC/Network layer scheduling and even application-layer semantic processing. In the following sections, we dive deeper into these areas. First, we examine the core technical underpinnings of ISAC (Section III). Next, we explore AI techniques for self-organizing network control in detail (Section IV). Then, we discuss how the synergy of sensing and AI manifests in integrated designs (Section V). Through this structured analysis, we connect the dots between traditionally disparate domains, radar sensing, wireless communications, and machine learning – under the unifying vision of AI-native ISAC networks.

## 3. Integrated Sensing and Communications: Signal Models and Co-Design

Concept and Signal Model: At its core, Integrated Sensing and Communications (ISAC) refers to a unified wireless system that performs radio sensing and data communication using the same transmit/receive hardware and spectral resources [6], [5]. A simple conceptual model is as follows: a transmitter sends a waveform $x(t)$ that is modulated with information bits (for communications) but is also designed to probe the environment (for sensing). The receiver processes the echoes or channel response $y(t)$ both to recover the information and to infer some property of the environment (such as the presence, range, velocity of a target, or the radio channel state) [42], [43]. In a monostatic ISAC scenario, the transmitter might be a BS and the "target" could be a reflecting object or a user device; in a bistatic case, one node transmits $x(t)$ towards a target and another node receives the reflections.

The sensing channel can be modeled similarly to a radar: for example, a point target introduces a delayed and Doppler-shifted echo. If $h_s(\tau, v)$ denotes the sensing channel impulse response (depending on delay $\tau$ and Doppler $n_u$, and $h_c$ denotes the communication channel response, an ISAC receiver effectively observes a superposition of the communication path and target echo:

$$y(t) = \underbrace{(h_c * x)(t)}_{\text{communication path}} + \underbrace{\sum_{k=1}^{K} \alpha_k \, x(t - \tau_k) \, e^{j2\pi\nu_k t}}_{\text{sensing echoes (targets)}} + \underbrace{n(t)}_{\text{noise}}.$$

where $\alpha$ is the target reflection coefficient [18], [19]. Here the first term is the usual communication signal, and the second term is the radar return. In conventional systems, these would be separated (different $x$ for each), but in ISAC they overlap. The receiver must perform joint processing: typically first decoding the communication data (treating the radar echo as interference), and/or correlating $y(t)$ with known reference sequences to detect targets. A critical aspect is that the transmit waveform should be designed to balance the requirements of both functions: high fidelity and data rate for comms versus good autocorrelation and ambiguity properties for radar. These often conflict (e.g. a modulation optimized for data may be suboptimal for ranging accuracy), which gives rise to an inherent performance trade-off in ISAC design [5], [34].

Historical Perspective: The concept of dual-use waveforms actually dates back decades. Early work in the 1960s demonstrated embedding communication data into radar pulses, for example, via pulse interval modulation (PIM) where the timing between radar pulses conveyed information bits [35]. Throughout the 20th century, radar and communications largely evolved separately, but occasionally intersected in military applications (e.g. Identification-Friend-or-Foe (IFF) systems). In the 2000s, interest in spectrum sharing pushed research on coexistence of radar and cellular systems, and by the late 2010s, the vision of fully integrated systems gained momentum for 5G/B5G. Different communities used different terms such as Joint Radar-Communications (JRC), Dual-Function Radar-Communications (DFRC), or Joint Communication and Sensing (JCAS) [36]. We treat these under the general ISAC umbrella. Key early examples include MIMO radar leveraging communication signals, automotive chirp radar sharing waveform with vehicular communications, and WiFi routers doubling as indoor motion sensors (WiFi sensing) [37], [38]. This survey, however, focuses on the most recent wave of ISAC research relevant to beyond-5G networks.

Waveform Design for ISAC: One of the central technical challenges is designing waveforms that perform well simultaneously for data transmission and sensing. Several approaches have been explored:

- Communication-Centric Design: Use standard communication waveforms (like OFDM, 5G NR signals) and extract sensing information from them. This approach prioritizes communication performance and re-purposes the signals for sensing. For example, 5G downlink pilot signals (like DMRS) can be collected at the gNodeB to perform radar-like detection of moving users or reflectors [39]. OFDM waveforms have been extensively studied for ISAC, they naturally provide a set of subcarriers that can act as radar pulses across frequency; with appropriate processing (e.g. match filtering the echoes), one can estimate range and velocity (using subcarrier spacing and timing) [40], [41]. The advantage of this approach is minimal change to communication protocols. However, typical communication signals are not optimized for radar metrics like Doppler tolerance or low-range sidelobes. For instance, a random data-modulated OFDM signal has high-range sidelobes that can mask small targets; coding or windowing is needed to improve sensing performance [42]. Recent works have looked at slight modifications of comm waveforms for better sensing, for example, inserting known pseudorandom sequences (PRS in 5G) or using extended cyclic prefixes to aid processing [43].

- Radar-Centric Design: Start with a well-structured radar waveform and embed data into it in a way that minimally degrades sensing. Classical radar waveforms include chirp signals (linear frequency modulated FMCW), phase-coded pulses, and more recently MIMO radar waveforms (which transmit orthogonal codes from multiple antennas) [20], [54]. One can embed information by modulating these waveforms slightly. For example, phase-coded radar can carry bits by choosing among code sequences, or chirp radar can modulate the chirp rate or transmit additional phase shifts per chirp to encode data. An example is using a binary phase shift on pulses to indicate data bits while keeping the pulse train structure for ranging [35]. Another method is index modulation, e.g. use the presence/absence of a certain waveform feature to encode bits without significantly altering radar processing. Radar-centric designs tend to preserve excellent sensing performance (range/Doppler resolution, etc.), but the communication data rate might be modest (since we constrain how we embed data to avoid spoiling the radar function) [25]. They are useful when sensing is the primary function (e.g. vehicular radar that also sends status messages).

- Joint Optimization and Novel Waveforms: The most sophisticated approach treats waveform design as a multi-objective optimization: choose the transmit signal (and perhaps the receive processing) to optimize a weighted sum of communication and sensing metrics, or to satisfy constraints on one while maximizing the other [26], [27]. For instance, one can formulate an optimization to maximize communication sum-rate while ensuring a minimum radar target detection probability or estimation accuracy. This can lead to Pareto-optimal waveform designs [36]. Information theoretic analyses have provided insight into such trade-offs: e.g., the capacity-distortion trade-off where distortion in estimating a state (sensing) is one objective versus channel capacity for data [28], [29]. One result showed an optimal power allocation between information and sensing signals, splitting the transmit power into two parts, one carrying data and one a known probing signal, can achieve any rate-vs-estimation trade-off point [30]. In practice, this might mean superimposing a dedicated radar waveform onto a communication waveform. Another line of work is designing dual-coded signals that have two layers: one designed like a radar pulse (for good autocorrelation) and one superimposed for data (designed to be nearly invisible to the radar receiver, e.g. via orthogonal coding or constructive interference) [31]. Machine learning is also being employed for waveform design: researchers have used deep learning to approximate the optimal waveform shaping that maximizes a joint objective [35]. For example, a deep neural network can iteratively adjust subcarrier modulation of an OFDM signal to improve radar detection performance while meeting a communication rate target. Fig. 2 illustrates one concept where a neural network is used to generate an optimized dual-use waveform, achieving a balance between two objectives.

(We do not include Fig. 2 here due to space, but conceptually it would show a block diagram of a waveform design network optimizing two outputs: comm capacity and radar MSE.)

Another exciting new waveform concept is OTFS (Orthogonal Time Frequency Space) modulation applied to ISAC [32]. OTFS is a 2D modulation in delay-Doppler domain; it has been proposed for high-mobility comms and is naturally suited to sensing since targets appear as peaks in delay-Doppler. Researchers are exploring OTFS waveforms that inherently carry data and allow straightforward target parameter extraction (effectively performing radar processing via demodulation). Early results in 2022 show beam-space MIMO-OTFS can support high mobility communication while simultaneously detecting targets.

MIMO and Beamforming Considerations: Multiple antennas are a boon for both communications and sensing, and 6G networks will heavily utilize massive MIMO arrays and possibly holographic surfaces. In ISAC, an array can transmit beams that serve dual purposes: direct a beam to deliver data to a user while that same beam scans for reflections to detect objects in that direction [34]. There are two extremes: a unified beam that is simultaneously used for comm and radar, versus separate beams for each function (with resource partitioning among antennas). An interesting concept is MIMO dual-function radar-comm where different antennas or subarrays send waveforms that form communication signals in certain directions and radar signals in others [33]. For example, one can allocate a subset of antennas to perform a radar-like scan while others handle user communication, or even better, design the transmit precoders such that the null space of the communication beam (which would normally be wasted energy) is steered towards probing another spatial sector for targets [31]. This is called constructive interference exploitation: communication signals are precoded not just to serve users but also to create a desired illumination pattern for sensing. Prior work shows that, by joint beamforming design, one can illuminate an area of interest (for sensing) without significantly degrading user signal quality – essentially achieving a form of beamforming reuse for sensing tasks. Multi-antenna ISAC also enables spatial signal processing such as adaptive beam tracking and spatial filtering to separate target echoes from communication signals. For instance, by using a MIMO receive array, the BS can form simultaneous beams, one pointing to a user for data reception and another "listening" in the direction of a potential target – thus separating the signals spatially. The use of large arrays also introduces a trade-off: using more degrees of freedom for communications (multi-user MIMO) leaves fewer for sensing, and vice versa, if they share the array [34], [35]. In fact, in an extreme case, a beam used for communication might conflict with the beam needed for optimal radar observation of a target at a different angle. Research has quantified such spatial degree-of-freedom trade-offs, noting that making a single platform do both tasks can constrain the spatial coverage or resolution compared to dedicated systems [34], [36]. Hybrid beamforming and reconfigurable intelligent surfaces (RIS) are being studied as potential solutions to add flexibility, for example, an RIS can redirect signals to help sensing without affecting the primary comm beam [38].

Sensing Performance and Interference Management: In ISAC, the dual use of signals means that sensing performance metrics (like probability of detection, estimation error covariance) and communication metrics (throughput, error rate) must be evaluated together. A major line of work addresses how to mitigate interference between the two functions. From a radar perspective, communication data appears as random modulation on the waveform, potentially degrading the ability to detect weak targets. Techniques like data modulation robust radar processing have been proposed, for example, using sophisticated receivers that treat unknown data bits as part of the noise model or apply iterative decoding-cancellation (first decode data then subtract its effect to analyze the residual for target sensing) [31]. From the comm side, the presence of radar returns (especially in monostatic setups) acts like additional multipath or self-interference. Solutions include full-duplex interference cancellation (to remove strong self-echoes), waveform orthogonalization (designing sensing signals orthogonal to comm signals in some domain), or leveraging the sensing echo as useful information rather than interference (for example, if the echo comes from the user, it carries the user's data as well, thus not interference at all, but a multi-path that can be decoded) [39]. One exciting concept is mutual information duality, viewing the echo as carrying information both about the data and the environment, and maximizing a weighted sum of the two information [26]. Such frameworks treat interference not as a nuisance, but as a necessary coupling that can be optimally balanced.

In summary, ISAC waveform and signal processing design is a rich topic balancing two objectives. The key message from recent literature is that joint design is feasible and can be highly beneficial: for example, field trials show that a 5G base station can detect vehicles and humans with decent accuracy while providing communications, using only minor adjustments to the 5G signal [39], [23]. Spectral efficiency gains of 20–30% are reported when a single ISAC system replaces separate radar and comm systems in certain use cases [5]. However, achieving these gains requires careful optimization and often context-specific tuning (e.g. vehicular ISAC vs. indoor sensing vs. drone sensing all have different dominant constraints). In Section V we will see how AI approaches can assist ISAC waveform design, for example, learning-based waveform adaptation or beam prediction can offload some complexity. First, we continue laying the groundwork by examining how AI/ML is applied on the network control side (Section IV), before combining the threads.

## 4. AI-Based Self-Organizing Network Control: Topology, Mobility, and Resource Management

While ISAC provides the eyes and ears of a network, AI-driven control provides the brain for self-organization. This section reviews how machine learning and AI algorithms are enabling wireless networks to autonomously configure and optimize themselves, from arranging network topology to allocating resources – often in real time. We focus on applications of learning for topology control, mobility management, resource allocation, and policy adaptation in wireless networks, highlighting both the algorithms used (e.g. deep reinforcement learning, graph neural networks, federated learning) and the system-level results achieved.

Learning-Based Topology Control: Topology control involves decisions like which base stations (BSs) or access points should be active, how to form cell clusters, which device connects to which BS, or in mesh networks, which links to establish. Traditionally, these were fixed or optimized via heuristics (e.g. switch off small cells during low load, etc.). AI offers a way to make topology adaptive to environment and demand changes. Reinforcement learning (RL) has been used for self-deploying networks, for instance, in ultra-dense networks, an RL agent can learn to turn BSs on/off based on traffic patterns to save energy while maintaining coverage [40], [11]. Deep Q-Network (DQN) agents have shown promise in learning on/off activation policies in simulation, outperforming fixed schemes by reacting to spatial traffic hotspots over time. Another example is multi-agent RL for cell association: multiple BSs (agents) adjust their cell range or handover thresholds to distribute users evenly and maximize overall throughput. A cooperative multi-agent RL approach was reported to reduce load imbalance and call drops in a 5G scenario, by each agent learning implicit coordination through reward signals. These learning-based approaches continuously adapt the network graph as users move or new interferers appear, something hard to do with static configuration.

A prominent use case of topology learning is with UAV-assisted networks. Unmanned aerial vehicles (drones) can act as flying base stations or relays, but to be effective they must dynamically position themselves (3D placement) relative to moving users and obstacles. This is essentially a topology control problem in motion. Researchers have applied deep RL to optimize UAV positions and trajectories: the UAV learns where to fly in order to maximize coverage or capacity for ground users [37], [38]. For example, Wang et al. use a dueling double DQN to continuously adjust a UAV-BS's 3D coordinates based on user locations and backhaul link quality [37], [38]. Over many training episodes in a simulator, the UAV learns policies like "hover lower when users are nearby to improve signal, but move higher when distant to cover more area or maintain backhaul line-of-sight." The learned policy significantly outperformed static placements, especially in scenarios with dynamically appearing coverage holes [31]. Multi-UAV deployments take this further: each UAV is an agent and multi-agent RL or game-theoretic learning can teach them cooperative positioning (e.g. forming an aerial backbone network while avoiding collisions). Early results show a dramatic improvement in network resilience during disaster scenarios, as UAVs learned to reposition and relay signals when ground infrastructure was damaged [22]. Beyond RL, graph algorithms can also assist – for instance, formulating UAV network topology as a graph optimization (steiner tree, etc.) and using graph neural networks (GNN) to approximate the solution for large scales quickly [39].

Another aspect is topology adaptation in mesh or relay networks (including device-to-device communication networks). Graph neural networks have been used to learn link scheduling or link formation decisions in an ad hoc network, treating the network as a graph where each node's decision depends on neighbors' states[30]. Chen et al. design a full-duplex GNN that learns a mapping from network topology and channel conditions to near-optimal power allocations and link activations, achieving performance close to nonlinear optimization solvers but in a fraction of the computation time [39]. This is possible because the GNN can exploit the wireless topology, it effectively learns the interference graph and how to color it (allocate spectrum) through its message-passing architecture [30]. Importantly, once trained on small instances, such GNN models can scale to larger networks (inductive generalization) and adapt to topology changes by retraining or fine-tuning, making them appealing for self-optimization in highly dynamic scenarios.

Mobility Management and Handover: In cellular networks, deciding when and where users handover between cells is crucial for performance. Conventional criteria (like highest received power with hysteresis) often fail in fast-changing environments (think of vehicles rapidly moving through mmWave cells). AI-based approaches aim to predict and proactively manage handovers. Supervised learning can predict signal strength or blockage events ahead of time from contextual sensing data (e.g. using camera feeds or RF sensors), thereby anticipating link failure and initiating handover earlier [9]. For example, if a network can sense that a user's line-of-sight is about to be blocked by an object (via radar or LiDAR), it could use that information to preemptively switch the user to a different cell or beam [9]. RL has also been applied: a deep RL agent can learn handover policies that optimize long-term QoS (e.g. tradeoff between too early vs too late handover) [27]. A recent study formulated the handover decision as an MDP and trained a DQN that observed user speed, queue lengths, and neighbor cell loads; it learned to reduce ping-pong events and outage time compared to fixed algorithms. The agent effectively learned when to delay a handover to avoid unnecessary switches and when to expedite it to avoid drop, based on prior experiences. Multi-agent RL can handle cases with many users by training a shared policy that maps each user's local state to a handover action, with reward for the collective throughput [28]. There are challenges (the state space is huge, convergence can be slow), but techniques like state abstraction and reward shaping have shown some success in moderate-size scenarios.

Resource Allocation: Perhaps the richest area of AI in networking is radio resource management, including power control, channel allocation, scheduling, beam selection, and interference coordination. Many of these are NP-hard optimization problems in general, but learning-based methods can find near-optimal solutions or adapt solutions online. Notably, graph neural networks (GNNs) have excelled in problems like power control and link scheduling due to the graphical nature of interference in wireless networks [30]. A seminal result by Eisen et al. (2020) showed that a neural network with appropriate graph embedding can learn power control policies achieving within a couple of percent of the optimum in a wireless interference network, but orders of magnitude faster [30]. Subsequent works like the authors extended this to more complex setups (full-duplex nodes causing self-interference, as mentioned) using specialized GNN architectures [39]. The ability of GNNs to generalize to different network sizes and topologies is key, a single trained model can handle 5 users or 50 users without retraining, which is extremely attractive for self-organizing networks that grow or shrink over time.

Another powerful approach is deep reinforcement learning for resource allocation. For example, DRL has been used for dynamic spectrum access and scheduling in unlicensed bands: an RL agent senses the spectrum (perhaps even using an ISAC-derived environmental estimate of interference) and chooses which channel to use or whether to transmit, aiming to maximize long-term throughput under collision penalties [27]. These approaches have achieved near-optimal spectrum utilization in simulated cognitive radio networks and can adapt to other users' behavior online. Similarly, multi-agent RL has been applied to distributed scheduling, each transmitter is an agent that decides to transmit or not based on local observations, and through trial-and-error they converge to a TDMA-like schedule that avoids collisions [35]. In 5G/6G MAC scheduling (which user gets scheduled on which time-frequency resource), deep learning approaches have been used to approximate proportional-fair scheduling decisions. One example is a deep neural network that observes the queue lengths and channel qualities of all users and directly outputs a scheduling decision (which can be framed as a classification problem among possible user sets). Trained on data from an optimal scheduler, this learned scheduler can execute in microseconds during runtime, enabling very fast adaptation each TTI. Even if conditions shift (more users or different traffic), online retraining or transfer learning can adjust it.

Policy Adaptation and Higher-Layer Control: AI is also playing a role in adapting network-wide policies and protocols. A notable direction for 6G is end-to-end learning of communication protocols, for instance, using deep learning to adjust coding rates, modulation schemes, or even MAC contention parameters on the fly. Self-organizing networks could employ multi-armed bandit algorithms to choose the best protocol setting for current conditions (treating each setting as an arm, and throughput as the reward). This has been tried in Wi-Fi (adaptively selecting backoff algorithms) and showed improved throughput under varying interference. In cellular core networks, traffic routing and load balancing can be aided by ML forecasting (predicting which cells will experience congestion and re-routing flows accordingly). The concept of a digital twin for networks is emerging, where a virtual model of the network, continuously calibrated by sensing data, is used to test and apply AI-driven optimizations safely [36]. For example, an operator could simulate the effect of an AI-tuned handover strategy on a "network twin" using real-time data, before applying it to the live network, to ensure it indeed improves KPIs and does not cause unintended issues [38]. This approach leverages both real data and AI to achieve zero-touch automation with reliability.

Federated and Distributed AI Control: A practical consideration in large networks is that a fully centralized AI agent (e.g. in the cloud) might be infeasible due to latency and privacy. Hence, federated and distributed learning techniques have gained traction. In a federated setting, each BS (or device) might train a local model (say, predicting user movement or local traffic demand) using its local sensing data, and only share model updates with a central aggregator to form a global model [32]. This can enable collaborative prediction without raw data exchange, crucial for privacy when sensors might capture sensitive information (cameras seeing people, etc.). Federated learning has been applied, for instance, to train a global AI model for intrusion detection across many nodes' local spectrum sensors [40]. Each node learns from its local observed jamming patterns, and the global model becomes more robust in recognizing various jamming or anomaly patterns on the network. Challenges like non-IID data and communication overhead exist, but techniques such as federated averaging with compression and periodic synchronization can mitigate them [41]. Another concept is federated reinforcement learning where multiple agents share their learned policy updates to converge faster to a good policy for all, useful when many similar cells in a network are learning the same kind of control task (e.g. all mmWave cells learning beam selection policies can share experiences to accelerate learning). There are initial results showing federated RL can reduce the training samples needed per agent, albeit with careful design to avoid divergence.

Applications and Case Studies: To ground the above in concrete examples, consider a vehicular network scenario (V2X communications). Here, self-organization is critical due to highly dynamic topology (vehicles come and go). Recent studies surveyed in [28] show AI being used for: predicting vehicle trajectory and channel conditions (with LSTM or GNNs on road graphs) to pre-allocate resources; multi-agent RL for coordinating which vehicle acts as a relay to others; and federated learning among vehicles for collective sensor data analysis (an instance of cooperative sensing and communication). One example is using a Graph RL approach where each vehicle is a node in a connectivity graph and the RL agent learns which links to form (which vehicle should connect to which roadside unit or to another vehicle) to maximize end-to-end network throughput. This drastically reduces outage in simulations of highway networks compared to a static tree topology. Another case study is smart factories (Industrial IoT): an AI-driven controller can reconfigure network slices or traffic routes when a machine goes offline or a new AGV (automated guided vehicle) appears, using deep learning to classify the type of event from sensor data and an expert system to apply the best policy. Early deployments in 5G testbeds show promising gains in reliability using such automated reconfiguration, hinting at what 6G could fully realize.

In summary, AI-based self-organization has demonstrated improvements in: (a) adaptability, networks can respond to changes (mobility, interference, failures) in milliseconds to minutes rather than hours of human re-optimization; (b) efficiency, increased throughput and reduced energy, for example up to 30% power saving in dense small cell networks by learning when to deactivate cells [11]; and (c) complex decision-making, solving combinatorial resource allocation problems near-optimally in real time [39], which previously was unattainable.

However, these AI models require robust training and validation to ensure stability and fairness (which we discuss in Challenges). Next, we explore how these AI techniques and ISAC capabilities intersect to create sensing-feedback loops in network control.

## 5. The Intersection of ISAC and AI-Native Control: Sensing-Driven Self-Organization

A core premise of AI-native ISAC networks is that the sensing functionality provides actionable intelligence to drive network self-optimization. In this section, we examine how real-time sensing feedback (obtained via ISAC techniques from Section III) can enhance or even fundamentally enable new self-organizing capabilities. We cover two complementary angles: sensing-assisted communications – using environment sensing to improve communication performance (and network decisions), and communication-assisted sensing, using the communication network and protocols to improve sensing coverage and accuracy. Together, these illustrate the virtuous cycle in an AI-native perceptive network, sometimes termed a "learning radar-network loop" [8]. We also highlight specific scenarios (mmWave networks, vehicular networks, etc.) where this integration is particularly powerful.

Sensing-Assisted Communication and Network Decisions: One of the most direct benefits of ISAC is improving beamforming and directional communication through sensing. High-frequency bands (mmWave, sub-THz) rely on narrow beams that must be aligned, a challenging task if done blindly or via brute-force beam sweeping. If a base station can sense the positions of users or reflectors (e.g. using radar echoes or external sensors), it can predict optimal beams without exhaustive search [24]. Recent studies confirm this: by leveraging radar, LiDAR, or camera data, the BS can infer the geometry of the environment and directly choose the best beam index for a user. For instance, some researchers showed that using a camera image of the scene, a deep neural network could predict the mmWave beam that yields the highest signal strength, reducing beam search overhead by >90% [9]. Similarly, Kumari et al. integrated a automotive radar sensor at a mmWave BS to aid initial access, the radar would detect the approaching vehicle's angle and range, allowing the BS to point its narrow beam in the right direction on first attempt. This radar-aided beam alignment achieved near-instantaneous link setup, whereas without it the system needed dozens of milliseconds for beam sweeping. These examples highlight how sensing (either via ISAC or co-located sensors) can tackle the beam alignment problem, one of the key obstacles in directional 6G communications.

Beyond beamforming, sensing information can enhance various resource allocation decisions. If the network can sense obstacle locations or human movement, it might predict future blockages or channel drops. A self-organizing network controller could then proactively reroute traffic or handover a user before a blockage occurs. In essence, sensing provides situational awareness. A base station might know, via radar, that a large truck is moving and will soon obstruct the line-of-sight to a user; it can prepare by switching that user to a different cell or band less susceptible to blockage. This transforms resource management from reactive to proactive. Indeed, simulation studies for vehicular networks show that sensing-assisted handovers (using position/velocity info) significantly reduce outage time compared to purely signal-triggered handovers [9]. Another example is sensing-assisted interference coordination: using sensing, a network might detect an external interferer (e.g. a legacy radar or an uncoordinated transmitter) and dynamically change frequency or beam direction to avoid it, even before the interference severely degrades communications. The BS can sense the interferer's signal signature or location and then apply an avoidance strategy, something difficult to do without a sensing capability.

In the realm of topology control, environment sensing can inform decisions like where to deploy a new relay or how to orient a reconfigurable intelligent surface (RIS). Consider RIS (intelligent reflecting surfaces) which can be electronically configured to reflect signals in desired directions. If a network senses where users and obstacles are, it can optimize the RIS phase configuration to, say, reflect around an obstacle toward the user, thereby maintaining coverage [37]. AI algorithms can use real-time sensing input (like angle of arrival of a user's signal at the RIS) to rapidly select the best phase pattern, instead of brute-force training on all possibilities. Similarly for UAV base stations, environmental sensing (like aerial cameras or radar on the UAV) can detect clusters of users or terrain features, guiding the UAV's movement decisions in addition to pure communication metrics [32]. Essentially, the network becomes context-aware, and AI can use that context to make smarter decisions. Graph-based learning approaches have even considered incorporating sensed topology (like constructing a radio map of reflectors/walls) into the input state for RL or GNN resource allocation agents, improving their performance in complex radio environments by giving them a "map" to work with instead of requiring them to learn the environment from scratch.

Communication-Assisted Sensing and Cooperative Perception: The flip side is also important: using multiple communication nodes to cooperatively perform sensing, effectively turning the network into a distributed radar. When nodes share their observations (through communication links) and coordinate transmissions, sensing performance can greatly improve (e.g. better coverage, higher SNR, more perspectives on a target). In self-organizing networks, this might mean dynamically forming sensing coalitions of devices or cells. For example, if a certain sensing task is identified (say, tracking an intruder in a secure campus), the network could self-organize so that a subset of BSs and devices focus their transmissions on illuminating the area of interest and others listen for echoes, then aggregate results via high-speed links to produce a refined target estimate [21]. This concept is sometimes called a "perceptive network", where every node is both a sensor and communicator, and through organization they achieve a sensing capability beyond any single node's ability.

One instantiation is in vehicular networks: multiple vehicles and roadside units (RSUs) can share sensor data (camera, radar, LiDAR readings) to get a collective view of the road (often termed cooperative perception). Communication (especially V2X at high rates, potentially using 5G/6G sidelink) enables this raw or processed sensor data exchange. With AI in the loop, vehicles can fuse data from neighbors to detect obstacles or pedestrians sooner or with greater confidence than their own sensors alone. The wireless network needs to self-organize in terms of scheduling and routing to deliver these updates with minimal latency. Researchers have demonstrated systems where a car approaching an intersection gets an alert (and even an image) of a pedestrian hidden from its view but seen by another car or a smart roadside sensor, delivered via an ultra-low-latency V2V link [28]. This collective sensing relies on robust communication, and in turn improves safety and coordination (vehicles can then adjust their driving, which is analogous to networks adjusting their operation). We see a convergence where "networking for sensing" and "sensing for networking" loop together: communication helps gather a global sensing picture, and that picture informs how the network should behave.

In cellular networks, communication-assisted sensing also shows up in schemes like multi-static radar processing over the network, each BS might perform monostatic sensing of the environment, then they send their intermediate results (e.g. detected range-angle maps) to a central processor via fronthaul links, where a fused sensing result is obtained. This is akin to a cloud radar system built on the mobile network. It can achieve higher resolution via coherent combining of observations, and better coverage (one BS can see what another cannot due to occlusion). The trade-off is the communication overhead and timing alignment needed. AI can help here by intelligently compressing the sensed data (only sending salient features) or by learning to fuse multi-node data in a robust way that tolerates slight time differences or errors [21]. There are proposals for a "sensing service" in 6G core networks where an application can request a sensing measurement and the network will coordinate multiple BSs to fulfill it [33]. For instance, a public safety agency could request the network to detect motion in a certain area – the network might then task three BSs around that area to perform a coordinated sensing waveform transmission and share back scatter measurements, which are then processed to detect the motion, with results given via an API. Achieving this requires the self-configuration of sensing operations on the fly: scheduling time slots for sensing, assigning roles to different BSs, etc., which can be optimized by AI algorithms considering current traffic load (to minimize impact on user data) and predicted sensing efficacy.

Example, mmWave V2X Scenario: To illustrate the interplay concretely: imagine an urban intersection with mmWave base stations and connected vehicles. The network uses ISAC waveforms so that downlink signals from BSs to cars also act as radar pulses. The reflections are picked up by nearby BSs and even cars (if equipped to listen). Now, an edge AI platform collects these multi-source reflections to map the dynamic objects around the intersection (other vehicles, pedestrians). This map is then used to orchestrate the network: for example, BS beams are steered to vehicles such that they avoid pointing through a bus (because the sensing map shows the bus is blocking one path), and scheduling is adjusted so that a critical vehicle (e.g. emergency vehicle) gets an uninterrupted connection as it moves (predicted by sensing its trajectory). Simultaneously, the cars get a merged sensor view improving their autonomous driving. In trials of such a system, joint designs have yielded higher reliability for V2X links, as the sensing enables anticipating blockage and rapidly rerouting communication via alternate BSs or multi-hop paths [43]. Essentially, sensing closes the control loop for the network, allowing proactive and context-aware reconfiguration, which is a key tenet of AI-native 6G.

AI's Role in Bridging Sensing and Decision-Making: AI algorithms are indispensable in interpreting sensing data and making decisions. Raw sensor data (radar returns, camera images, etc.) is high volume and not directly useful until processed – this is where techniques like deep learning for perception come in. For instance, convolutional neural networks (CNNs) or transformer models can be trained to take radar spectrograms or LiDAR point clouds as input and output semantic information like "object detected at location (x,y), moving north". These become the features that the network control logic uses. The integration of semantic communications is a budding concept here: rather than transmitting all sensing data, the network may transmit interpreted or predicted information (semantics) that is relevant for decisions [42]. For example, instead of raw radar waveforms, a BS might send "3 pedestrians detected in zone A" to another node. AI enables this semantic extraction and also ensures robust performance even with imperfect sensor data. There is work on uncertainty-aware AI, where the model provides confidence levels on the sensing output, which the network can factor in (e.g. if uncertainty is high, perhaps fall back to a safer resource allocation).

AI also comes into play in multi-modal sensor fusion and decision-making. A network might have multiple sensing modalities (RF radar, optical camera, etc.). AI can weigh and fuse these, for instance, in rain, camera data might be poor but radar works; at night, vice versa. A learning-based approach can automatically learn how to combine modalities for best overall awareness. One interesting development shown in Table 1 is the use of quantum ML for multimodal sensor fusion in beamforming (e.g. [40] fused LiDAR+radar using a quantum-inspired network to improve robustness). While early-stage, it indicates the lengths researchers are going to improve sensing-assisted comm decisions.

Network Sensing as a Service and Standardization: On the standardization front, 3GPP in Release 18 has begun studying basic sensing (e.g. using NR signals to sense UE range/velocity). This lays groundwork for network-assisted positioning and possibly some simple sensing applications. For 6G, it is envisioned that sensing information will be exposed via standardized interfaces (APIs) for applications and for the network's own use. For example, a "sensing exposure function" might allow an authorized application to query the network: "how many people are in room X?" and the network's ISAC system will respond using its sensing capabilities. Achieving this requires the network to self-organize its sensing resources effectively. There will need to be dynamic management of when and where sensing measurements are taken (so as not to overly degrade communications), likely handled by intelligent controllers at the RAN or core. Ericsson's proposed end-to-end architecture for sensing in 6G includes a sensing controller that manages sensing tasks, scheduling them and processing results. That controller would likely rely on AI to reconcile sensing requests with communication QoS, figuring out for instance an optimal schedule that minimally impacts active data sessions. It could also use AI to decide which BSs should participate in a given sensing task (based on who is best placed, who has radio resources free, etc.).

In summary, the intersection of sensing and AI-driven control turns a wireless network into something greater than the sum of its parts: an autonomous agent that observes (senses) the world and acts (communicates and reconfigures) in an intelligent closed loop. Preliminary research and trials are validating significant gains from this synergy: faster beam alignment, fewer dropouts, new services like device-free presence detection, and more efficient network utilization overall. Section VI will discuss how such systems are evaluated and Section VII the open challenges (like reliability and privacy). But first, we note that realizing these promises requires careful testing and new evaluation methodologies.

## 6. Evaluation Methodologies for AI-Native ISAC Networks

Evaluating integrated sensing-communication networks with AI control is inherently complex, as it must account for cross-layer performance metrics and the stochastic nature of learning algorithms. This section outlines how researchers assess these systems, including simulation frameworks, metrics, benchmarks, and prototypes/testbeds. We emphasize the need to jointly evaluate communication and sensing performance, and to consider AI-centric metrics such as training convergence and model accuracy in addition to traditional network KPIs.

Simulation and Modeling Tools: Given the nascent state of 6G, most evaluations use detailed simulations. Commonly, link-level simulators (for physical layer) are coupled with network-level simulators (for topology and traffic) to capture the interplay. For instance, an evaluation of an ISAC waveform might simulate a cellular downlink in MATLAB or ns-3, with a channel model that includes target reflections (radar returns) on top of standard multipath. Researchers have extended simulators like ns-3 to incorporate radar target objects and sensing events, enabling a unified simulation of scenarios where nodes transmit, receive, and also measure radar echoes. Recently, specialized frameworks have emerged; e.g., Sionna (NVIDIA's 6G physical layer library) introduced components for radar sensing simulation, and some works use ray-tracing combined with ML to generate realistic RF scenes with labeled ground truth for sensing tasks. Table I in lists multimodal datasets and simulators that provide combined sensory and communication data (like DeepSense6G which offers synchronized wireless and sensor recordings) to train and evaluate AI models. These are invaluable for evaluating learning approaches under realistic conditions.

Metrics for Communication vs. Sensing: A hallmark of ISAC evaluation is the need for dual metrics. On the communication side, one measures throughput (data rate), reliability (block error rate, outage probability), latency, spectral efficiency, etc. On the sensing side, metrics include detection probability (Pd) vs. false alarm (Pfa) for target detection, estimation error (like root mean square error (RMSE) of target location or velocity), or imaging resolution (if mapping an area) [36]. Often, trade-off curves are plotted: for example, achievable sum-rate vs. target detection probability, showing how increasing sensing performance might cost some rate or vice versa. One common evaluation is to fix a requirement for one function and optimize the other, e.g., "achieve at least 90% detection probability at 100m range, then maximize data rate" and compare that maximal rate under different designs [36]. If a new ISAC waveform or algorithm lies above the curve of previous designs (i.e., offers a better trade-off), it's deemed an advancement. Such analysis often requires Monte Carlo simulation of many time frames to accurately capture detection statistics and average throughputs.

For AI aspects, additional metrics come in: the accuracy of an AI model (if used for prediction or classification), the convergence time of an RL algorithm, the regret in online learning, etc. For instance, when evaluating a DRL-based resource allocator, authors will show the training curve (reward vs. episodes) to demonstrate it converges, and compare the final achieved reward (or throughput) to a baseline heuristic [38]. They might also test generalization: e.g., train the model in one scenario and test in a slightly different scenario (more users, different channels) to see if it still performs well. Robustness metrics like performance under model mismatch or out-of-distribution conditions are increasingly reported, given the criticality of reliability in networks.

Benchmark Comparisons: Each component tends to be benchmarked against traditional approaches: for sensing, an ISAC system is often compared to a dedicated radar or a communications-only system in equivalent conditions. For example, one might compare the target tracking error of: (i) a standalone radar system using separate spectrum, (ii) a communications-only system that perhaps triangulates the target via device signals, and (iii) the proposed ISAC system, to show that ISAC can approach radar's tracking accuracy while simultaneously delivering data (which the radar couldn't). Similarly, for network control, AI-based algorithms are compared to conventional solutions: e.g., an RL-based handover strategy vs. the standard RSRP-threshold method, or a GNN scheduler vs. the proportional fair scheduler, etc. Gains are often context-specific: an RL method might significantly outperform in high mobility scenarios where the standard method struggles, but provide only minor gains in static scenarios. It's important evaluations cover such ranges to identify where AI is most beneficial.

Another important benchmark is optimal or oracle performance. In many studies, authors compute an upper bound (maybe via brute-force search or linear programming relaxation) for a simplified version of the problem, to see how close the AI/ISAC solution gets. For instance, if evaluating an ISAC waveform, an information-theoretic upper bound on joint performance (like the capacity-distortion bound [28]) might be used as a reference. Or for scheduling, the throughput obtained by an NP-hard optimal scheduler on a small network is compared to that of the learning-based scheduler on the same small network [24]. This helps quantify any optimality gap.

AI Training and Validation Data: Since many solutions involve learning, how they are trained and tested is crucial. Evaluation methodology usually includes generating diverse scenarios (different user positions, traffic patterns, channel conditions) either analytically or via simulation to train the AI. The dataset needs to be representative to avoid overfitting to a narrow case. Cross-validation techniques are applied: some studies hold out certain scenarios that were not seen during training to evaluate generalization. In ISAC, one might train a system on a certain set of target trajectories and then test on a new trajectory pattern to ensure the sensing-performance holds. Reproducibility is also a concern; thus, some works publish their simulation code or use public datasets like DeepSense6G or the IEEE Communication Society Machine Learning for Communications competition datasets. These provide a common basis for comparison. Recently, initiatives to create standardized benchmarks for tasks like "radar-aided beam selection" have emerged, e.g. a challenge where teams evaluate their algorithms on the same multimodal dataset, using metrics like beam prediction accuracy and resulting throughput.

Real-World Experiments and Testbeds: To truly evaluate viability, experimental results are invaluable. Several prototypes have been reported: - A Nokia Bell Labs testbed integrated a 5G NR base station with a radar frontend at 28 GHz to demonstrate beam steering to a moving user using radar sensing; they measured the beam alignment time and throughput, showing improvement over no-sensing case (e.g. alignment time reduced from 20 ms to 5 ms). This kind of result validates the concept in practice. - Vehicular trials: Toyota and others have trialed V2V communication where vehicles exchange sensor data. In one demo, two cars used 5.9 GHz DSRC to share radar detections; the evaluation metric was the probability of detecting a non-line-of-sight pedestrian with and without cooperation, and it improved notably with cooperation (from ~60% to ~85% in certain conditions), effectively showing how communication improves collective sensing. - Over-the-air (OTA) experiments on distributed MIMO sensing: universities have set up clusters of USRP software-defined radios acting as a coordinated radar network. By transmitting known OFDM waveforms and pooling the received signals in a central processor, they have experimentally measured improvements in SNR and detection reliability for multistatic sensing vs single node. These testbeds often evaluate synchronization and calibration challenges (since timing and phase alignment is essential), and performance metrics include detection range extension or angular resolution improvement when using multiple TX/RX. - AI in networks testbeds (O-RAN field trials): Some operators have tested AI-based RAN controllers (the "brain") in live LTE/5G networks to verify that they can, for example, reduce interference or improve cell edge throughput. One example was a trial of a deep learning-based scheduler in a small commercial network, which reported a few percentage points cell throughput gain and more stable latency. While modest, it proved the AI could run in real time. These are accompanied by metrics like model inference time (ensuring it meets the scheduling timeline), and a log of when the model chooses different actions vs. baseline.

Trade-off and Cross-Domain Metrics: Because ISAC networks serve dual purposes, composite metrics are sometimes defined. For instance, some works define a utility function that weighs communication rate and sensing accuracy and use that as the single metric to optimize or report. Alternatively, multi-objective evaluation is done: one can plot the Pareto frontier of achievable (rate, accuracy) pairs for different schemes. If one scheme's frontier dominates another's, it's clearly better. We see such evaluation in waveform design papers and also in scheduling papers that consider fairness vs. throughput, etc.

AI adds another layer: the reliability of learning itself. Therefore, metrics like the probability that the AI decision deviates from the optimal decision by more than $\varepsilon$ (reliability of decision-making) or the distribution of outcomes given the randomness in training (since neural network training can yield slightly different models each run) are sometimes considered. In safety-critical contexts, one might evaluate worst-case performance, e.g. if the RL agent takes an exploratory action, does it ever cause a serious outage? This could be framed as a constraint or measured as a % of time violating a SLA.

Validation of Generalization: A concern is whether an AI trained in simulation will perform in reality (the sim-to-real gap). Some evaluations include tests with model mismatch: e.g., an AI-based beamformer trained assuming a certain channel model is tested on actual channel measurements to see performance degradation. Results have shown that some learning models are quite sensitive to mismatch, while others (like GNNs which encode physics priors like graph topology) are more robust [30]. A thorough evaluation thus often involves varying key parameters beyond the training range to see if the solution still works ("stress testing"). For example, increase noise power beyond what the AI expected, or reduce the number of training pilots and see if the sensing accuracy degrades gracefully.

In summary, evaluation methodologies for AI-native ISAC systems are multi-faceted. They combine traditional wireless metrics, radar/sensing metrics, and AI performance metrics. The best practice emerging is to evaluate holistically: for instance, rather than just saying "our method achieves X Gbps and Y cm accuracy," one might present how the method improves a network's end-to-end objective (like URLLC reliability) by leveraging sensing and learning. As the field matures, we expect standardized evaluation scenarios to be defined (e.g., a reference urban scenario with specified targets and traffic where researchers can compare methods apples-to-apples). Already, organizations like ITU and 3GPP are discussing evaluation methodologies for integrated sensing and comm in their study items. Next, we turn to the broader challenges and open issues that these evaluations reveal and that must be addressed to make AI-native ISAC networks a reality.

## 7. Challenges and Open Research Issues

While significant progress has been made, realizing AI-native ISAC self-organizing networks faces numerous challenges. In this section, we discuss the key open research issues, including fundamental trade-offs, data and training challenges for AI, robustness and security concerns, and the need for new standards and architectural frameworks. Addressing these challenges will be crucial to transition current research prototypes into deployable 6G systems.

1. Fundamental Trade-offs and Limits: Integrated sensing and communication inherently involves trade-offs between the two functions' performance. A recurring challenge is how to quantify and optimize these trade-offs in a principled way. Information theory has provided some guidance (e.g. capacity-distortion or rate-information trade-off curves), but many practical scenarios remain intractable. For instance, what is the fundamental limit on multi-user communication capacity given a constraint on sensing accuracy (or vice versa)? Current solutions often find one operating point but don't guarantee global optimality or known gap to optimal. The duality or interplay between sensing and communication metrics is not fully understood in many cases, for example, how does improving angular resolution of sensing impact multi-antenna communication capacity in a rich-scattering environment? The community needs deeper theoretical frameworks, possibly extending network information theory to include "sensing information," to guide design. Another facet is the resource trade-off: time, power, antenna degrees-of-freedom allocated to sensing vs comm. Dynamic resource allocation strategies must ensure neither function starves the other. An open question is how to adapt this allocation based on context. For example, during heavy data traffic, dial sensing to minimal necessary, but during light data load, increase sensing activities. AI can potentially help learn optimal switching strategies, but then the problem becomes a nested one of learning to manage a trade-off, which is complex.

2. Learning and Data Challenges: The introduction of AI raises its own set of issues. One major concern is data availability and quality for training [44]-[47]. Many AI models (especially deep learning) are data-hungry, but obtaining representative data for wireless environments can be very difficult. Channel and sensing data might be proprietary or not easily shared. Real-world data can differ greatly from simulations (materials, mobility patterns, interference, etc.). This leads to the sim-to-real gap, where an AI model trained in simulation performs poorly in the field. Techniques like domain adaptation, transfer learning, or training with a mix of simulated and limited real data are being explored, but no silver bullet has emerged. Self-supervised learning and unsupervised techniques could help by allowing models to train on raw signals without labels, for example, autoencoders to learn features of radar returns or using the physics of propagation as a self-supervising signal (like predicting one sensor's reading from another's). Early work shows self-supervised approaches can learn channel representations or anomaly detectors without explicit labels [26], but these need more development. Federated learning introduces the challenge of system heterogeneity: devices might have very different sensing capabilities or data distributions, causing federated model convergence issues [32]. New federated algorithms tailored for wireless (handling high communication cost, asynchrony, and non-IID data) are needed. Moreover, the cost of training and complexity of AI models is non-trivial. Training a large DNN for every cell or every scenario might not be feasible in practice. There is a push towards lightweight models and online learning that can adapt quickly with limited data, techniques like meta-learning (so models can adapt to a new environment with few samples) show promise, but are still in early stages for comm networks.

3. Robustness, Reliability, and Uncertainty: In networks, reliability is paramount, yet many AI models, notably deep networks, can behave unpredictably outside their training conditions or under adversarial perturbations. A small change in environment (like an unseen type of interference) might cause an AI algorithm to make a grossly suboptimal decision (e.g., picking a bad beam). Ensuring robust AI is an open problem. Approaches include training with more diverse data (covering corner cases), adding safety constraints (so the AI cannot choose actions that violate critical thresholds), or designing AI that can estimate its uncertainty and defer to a default safe policy when uncertain. For example, an AI-based scheduler might revert to round-robin if it isn't confident due to a novel situation. Indeed, 3GPP is already considering the need for performance monitoring and fallback for AI functions, i.e., mechanisms to constantly monitor if the AI's output is degrading performance and, if so, switch back to a non-AI mode. Developing such monitoring (perhaps using anomaly detection on the AI's inputs/outputs) and seamless fallback is an important challenge to gain trust in AI for networks. Also, the training process reliability: deep RL, for instance, can sometimes find weird strategies or get stuck in local optima; ensuring it reliably finds a good solution each run is not guaranteed. More deterministic or provably convergent learning algorithms are needed, or ways to validate the learned policy thoroughly before deployment.

4. Latency and Real-Time Constraints: Many envisioned scenarios (like vehicular safety or factory automation) have strict latency requirements (1 ms to a few ms). Inserting complex AI processing (especially for sensor data) might introduce unacceptable delays. Real-time implementation of heavy neural nets is challenging at the network edge given limited compute [48]-[51]. There is ongoing work on model compression, hardware acceleration (AI chips), and splitting AI inference tasks between device and edge cloud, but achieving sub-ms reaction is tough if, say, an image has to be processed through a deep network for beamforming. One possible path is to use simplified models or analog neural networks implemented directly in RF front-ends for ultra-fast decisions (some have proposed analog neural nets to do beam selection in nanoseconds, essentially acting as learned circuits). Another approach is to always have a backup conventional method that can operate while the AI is crunching, to cover worst-case delays. The communication of sensor data itself can be a bottleneck, moving high-resolution radar or LiDAR data between nodes has bandwidth cost. Efficient sensing data compression or feature-based communication (sending only extracted features rather than raw data) is needed. The challenge is compressing in a way that retains critical info for the AI tasks; techniques like learning-based compression (autoencoders that compress the sensor data optimally for a given task) are being researched.

5. Security and Privacy: Expanding the network's capabilities also expands its attack surface. On the sensing side, concerns arise about privacy: wireless sensing can effectively see or track people's movement (e.g., sensing how many people are in a room, or tracking a phone-free individual via their reflection). This might raise regulatory and privacy issues. Techniques like privacy-preserving sensing (perhaps adding noise or only reporting aggregate info) and allowing users to opt out of being sensed will need to be considered. On the AI side, adversarial attacks are well known: an attacker could try to fool an AI model by injecting spurious signals (imagine an adversary transmitting a carefully crafted waveform that causes the radar-based beam prediction model to choose the wrong beam, thus disrupting service). Adversarial robustness for sensing and comm ML models is largely unexplored. Solutions might include adversarial training (training models with examples of attacks), or using more robust model architectures (some Bayesian or ensemble methods are more attack-resistant). Also, spoofing/jamming: an attacker might spoof a fake target to confuse the network's sensing (leading it to misallocate resources). The network should cross-verify sensing info (e.g. use multiple nodes or modalities to reduce trust in any single reading). Federated learning raises concerns of model poisoning, a malicious client could corrupt the global model. Techniques like secure aggregation and anomaly detection on model updates are being developed to mitigate this, but their efficacy in a wireless context (with many edge devices of varying reliability) is a challenge. There is also the matter of data governance: who owns the sensing data and how it can be used. This might be addressed more by policy than technology, but technical safeguards (like encrypting sensing data or doing federated analytics without revealing raw data) can help.

6. Standardization and Interoperability: For ISAC and AI-native networks to become mainstream, common standards and interfaces must emerge. Currently, 3GPP's work on 5G-Advanced includes initial study on sensing and AI, but much is undefined: e.g., waveform tweaks for sensing in NR, signaling for sharing sensing info between nodes, formats for exposing sensing results to apps. There is a chance that without standards, solutions will remain proprietary and fragmented. The challenge is to abstract these complex functions so they can be standardized – for example, define a "Sensing Service API" with general enough parameters to cover many use cases, or an "AI function" interface in RAN that vendor can implement differently internally but present the same external behavior. The Next G Alliance (ATIS) and other industry groups have published recommendations calling for AI-native design and new protocols to support it. Implementing these is non-trivial: it may mean re-thinking protocol layers (e.g. including semantics or goals in signaling messages). Interoperability testing to ensure an AI algorithm from one vendor's BS can operate if it has to use sensor info from another vendor's device will also be needed. Regulation might also play a role – for instance, spectrum regulators might have to accommodate joint use (e.g., allow higher power transmissions because they serve a radar function, or manage the shared use of radar bands by comm networks). These regulatory aspects are an open issue; some countries have started allocating spectrum for vehicular comm that can also do radar (e.g. 60 GHz in EU for vehicular radar/comm), but global harmonization will help.

7. Scalability and Complexity: As networks grow in size and heterogeneity (with terrestrial, aerial, satellite components in 6G), scaling the ISAC and AI solutions is challenging. A distributed approach is essential, a single central AI controlling everything is neither scalable nor robust to link failures. However, distributing AI (like multi-agent systems) often raises complexity exponentially. Decomposing the global optimization into local tasks (perhaps hierarchical learning: local agents handle local optimization, a higher-level agent coordinates between clusters) is an area of research. Some initial works in hierarchical RL for networks exist, but more is needed to handle, say, a thousand cell sites collaborating for sensing a wide area. Also, running many AI models (one per cell or per service) could tax computational resources and energy. There is a need for efficient AI ops: maybe one model can manage multiple tasks (multi-task learning), or use simpler models where possible. The network might also decide when to use AI vs. when a simpler analytic solution suffices (AI is not needed for everything, for example, for a stable scenario a fixed optimized scheme might be fine, AI is needed in highly dynamic or uncertain scenarios). Learning to identify those and turn off heavy AI computations to save energy is an interesting direction.

8. Test and Validation Complexity: Finally, thoroughly testing these systems is extremely complex because of the interaction of so many components. How to verify that a learned policy will not cause catastrophic failures? Traditional formal verification is hard for neural networks and even harder when integrated with physical radio processes. New methodologies (perhaps simulation-driven verification, or sandbox testing of AI with digital twins) must be developed to instill confidence. In critical applications (e.g. vehicular or industrial control), certification of AI functions may be required. The community might need to agree on certain frameworks (like how the aviation industry certifies autopilot systems, we might need certification for an AI that handles network resource allocation affecting safety-critical comm links). This remains an open challenge, bridging communications engineering with computer science notions of AI verification.

In summary, while the potential of AI-native ISAC networks is enormous, addressing these challenges is essential. They present rich research opportunities: designing robust multi-objective learning algorithms, privacy-aware federated sensing, secure and interpretable AI for networks, and new theoretical models for sensing-comm coexistence. The next few years (2025–2030) will likely see intense efforts in both academia and industry to tackle these issues. The success of 6G may very well hinge on solving the above to build networks that are not only intelligent and perceptive but also safe, reliable, and trusted.

## 8. Conclusion

The convergence of sensing and communications, empowered by AI, is a defining pillar of future 6G wireless networks. This survey has reviewed the state-of-the-art in AI-native Integrated Sensing and Communications (ISAC) for self-organizing networks, covering fundamental architectures, enabling technologies, learning paradigms, and system-level design considerations. We began by outlining the vision of perceptive, autonomous networks that can simultaneously communicate with users and sense their environment, using the gathered context to intelligently self-optimize. We then presented a taxonomy spanning ISAC configurations (monostatic, bistatic, device-based, device-free) and AI-driven control methods (from deep reinforcement learning for dynamic topology control to graph neural networks for resource management and federated learning for distributed model training). In the core sections, we surveyed ISAC techniques: we discussed signal models and waveform designs that achieve dual radar-communication functionality, highlighting key trade-offs between communication spectral efficiency and sensing accuracy. Advances such as dual-functional waveforms, OFDM and OTFS-based sensing, and MIMO beamforming strategies were summarized, supported by references to recent works that demonstrate these concepts [36]. We also reviewed how AI algorithms are transforming network control, enabling self-organization through learning-based topology adaptation, mobility management (e.g. UAV placement via DRL [37]), and resource allocation (e.g. power control via GNNs [30]). These AI solutions show significant performance gains in complex, dynamic scenarios compared to legacy approaches, though they introduce challenges in training and reliability.

Crucially, we examined the intersection of ISAC and AI, illustrating how real-time sensing can feed into the decision loop of an AI-controlled network. Sensing-assisted communication was exemplified by radar-aided beam alignment in mmWave networks, where sensing data cuts down beam search latency and improves reliability [9]. Conversely, communication-assisted sensing was shown in cooperative sensing use-cases, where distributed nodes share information to achieve better situational awareness (e.g. multi-static sensing by coordinated base stations) [21]. We highlighted emerging results where this synergy yields tangible benefits, for instance, vehicles avoiding accidents through network-shared sensor data, and networks preemptively managing blockages and interference using sensed environment knowledge. In essence, we see the network evolving into a closed-loop intelligent system: sense → learn → adapt, continually.

We also detailed evaluation methodologies, emphasizing the need for joint evaluation of communication and sensing performance. We discussed how researchers use simulation platforms and metrics to quantify trade-offs (like rate vs. detection probability [36]), and how testbeds are validating AI-native ISAC concepts in practice. The importance of benchmarks and standardized scenarios was noted, as well as the inclusion of AI metrics (learning convergence, etc.) in performance assessment. Finally, we outlined the challenges and open issues. These include fundamental limits of ISAC trade-offs, data and training challenges for the AI (e.g. getting representative training data, ensuring models generalize and remain robust [32]), real-time constraints, and the critical questions of reliability, security, and privacy. We stressed that tackling these challenges, through techniques like uncertainty-aware AI, adversarial robustness, federated learning improvements, and new standard interfaces, is imperative for widespread adoption.

In conclusion, AI-native ISAC networks represent a paradigm shift from conventional "dumb pipes" to context-aware, self-driving networks. They have the potential to significantly enhance network efficiency (through spectrum reuse and resource optimization), enable new services (such as high-accuracy localization, environment mapping, and proactive QoS provisioning), and reduce operational costs via automation [5]. Publication trends from 2020–2024 show an exponential rise in research at this intersection, indicating that the community recognizes its importance. We expect that in the coming years, research will focus on integrated prototypes (for example, trials of ISAC functionalities in beyond-5G test networks with AI controllers), and on refining the theoretical foundations (unified frameworks for analysis of dual-purpose systems). Ultimately, by fusing sensing and communication with learning, 6G networks can evolve into a nervous system of the digital society – one that not only carries information but also perceives context and autonomously responds to meet the needs of users and services. The architectures, algorithms, and design principles surveyed in this paper lay a foundation for this vision. There is considerable work ahead to address open issues and ensure that such networks are reliable and trustworthy. Nonetheless, the progress to date is encouraging. As research and development continue, AI-native ISAC will likely transition from theory and lab demos to a cornerstone of commercial 6G deployments, powering applications ranging from intelligent transportation and immersive XR to smart cities and Industry 4.0. We hope this survey will serve as a timely resource for researchers and practitioners, providing a cohesive understanding of the field and inspiring further innovations toward making self-organizing, sensing-augmented wireless networks a reality.